
\magnification=1200
\pageno=1\tolerance=10000\hfuzz=5pt
\input harvmac
\hyphenation{Rob-er-ts}

\def\etal{{\it et al.}}

\def\pslash{p\!\!\!/}

\def\p2inf{\mathrel{\mathop{\sim}\limits_{\scriptscriptstyle
{p^2 \rightarrow \infty }}}}
\def\k2inf{\mathrel{\mathop{\sim}\limits_{\scriptscriptstyle
{k^2 \rightarrow \infty }}}}
\def\x2inf{\mathrel{\mathop{\sim}\limits_{\scriptscriptstyle
{x \rightarrow \infty }}}}
\def\Lam2inf{\mathrel{\mathop{\sim}\limits_{\scriptscriptstyle
{\Lambda \rightarrow \infty }}}}
\def\Q2inf{\mathrel{\mathop{\sim}\limits_{\scriptscriptstyle
{Q^2 \rightarrow \infty }}}}

\def\frac#1#2{{{#1}\over {#2}}}
\def\smallfrac#1#2{\hbox{$\frac{#1}{#2}$}}
\def\half{\smallfrac{1}{2}}
\def\third{\smallfrac{1}{3}}

\def\lsim{\mathrel{\rlap{\lower4pt\hbox{\hskip1pt$\sim$}}
    \raise1pt\hbox{$<$}}}         
\def\gsim{\mathrel{\rlap{\lower4pt\hbox{\hskip1pt$\sim$}}
    \raise1pt\hbox{$>$}}}         

\catcode`@=11 
\def\slash#1{\mathord{\mathpalette\c@ncel#1}}
 \def\c@ncel#1#2{\ooalign{$\hfil#1\mkern1mu/\hfil$\crcr$#1#2$}}
\def\lsim{\mathrel{\mathpalette\@versim<}}
\def\gsim{\mathrel{\mathpalette\@versim>}}
 \def\@versim#1#2{\lower0.2ex\vbox{\baselineskip\z@skip\lineskip\z@skip
       \lineskiplimit\z@\ialign{$\m@th#1\hfil##$\crcr#2\crcr\sim\crcr}}}
\catcode`@=12 

\def\matele#1#2#3{\langle {#1} \vert {#2} \vert {#3} \rangle }

\def\PR{{\it Phys.~Rev.~}}

\def\NP{{\it Nucl.~Phys.~}}
\def\PL{{\it Phys.~Lett.~}}
\def\PRep{{\it Phys.~Rep.~}}

\def\ZP{{\it Z.~Phys.~}}

\def\IJMP{{\it Int.~J.~Mod.~Phys.~}}

\def\vol#1{{\bf #1}}

\def\vpy#1#2#3{\vol{#1} (#3) #2}
\pageno=0
\vskip 1cm
\line{\hfill {\bf CERN-TH/95-71}}
\vskip 1cm
\line{\hfill {\bf DFTT 95/23}}
\vskip 2cm
\centerline{{\bf ANOMALOUS EVOLUTION EFFECTS ON }}
\centerline{{\bf SINGLET STRUCTURE FUNCTIONS}}

\vskip 1.5cm

\centerline{{\bf Marco Genovese}}
\vskip 1cm

\centerline{{\it Theory Division, CERN, CH-1211 Geneva 23, Switzerland }}

\centerline{{\it Dipartimento di Fisica Teorica, Universit\`a di Torino, }}

\centerline{{\it and INFN, Sezione di Torino, Via P.Giuria 1, I-10125 Turin,
Italy.}}

\vskip 1.5cm
\centerline{\bf Abstract}
\vskip 1cm
A study of the effects of the anomalous
evolution due to mesonic degrees of freedom on singlet structure functions
is presented.
Possible phenomenological applications are discussed.

\vskip 4cm

{\bf CERN-TH/95-71}
\vskip 0.5cm
{\bf March 1995}

\vfill\eject

One of the most successful results of perturbative QCD is surely
the detailed description of logarithmic scaling violations in
Deep Inelastic Scattering (DIS).
Recent improvements
in experimental accuracy are now opening up the possibility of
measuring subleading perturbative corrections, as well as corrections
to perturbation theory as expressed by higher twist effects.
However, it was pointed out  in \ref\balfor{R.~D.~ Ball and S.~Forte, \NP {\bf
B 425} (1994)
516.} that in the
region of $Q^2\sim 1$~(GeV/c)$^2$  the scale dependence of
the non-singlet nucleon structure function may be qualitatively rather
different from that predicted by purely perturbative QCD, due to
non-perturbative effects related to the anomalous breaking of axial
U(1) symmetry.

In \ref\BBFG{R.~D.~ Ball, V.~Barone, S.~Forte and M.~Genovese,
\PL {\bf B329} (1994) 505.} the effects on non-singlet structure functions
evolution were
studied, producing testable predictions of this model. This predictions are
in good qualitative agreement with the experimental data available
today and could
lead to a clear confirmation of the model, when more precise data will be
available.

Here we
will study the effects on the singlet structure function evolution,
which are expected in this scheme \foot{Incidentally, a previous tentative
qualitative estimate of these non-perturbative effects on singlet evolution
from the results obtained in the non-singlet case had led to
the suggestion \ref\df{S.~Forte and R.~Ball, \NP {\bf 39 B,C}
(Proc. Supl.) (1995) 28; QCD 94
Montpellier, June 1994 \semi M.~Genovese, 1994 PhD thesis, `Some aspects of
nucleon structure phenomenology: the deep inelastic scattering as a test bed of
QCD', Universit\`a di Torino in press.} that the effect could be opposite in
sign
and larger than the one we calculate here, giving a
possible explanation to the albeit marginal difference between the $\alpha_s$
extracted by DIS and LEP data \ref\PDB{See for example
`Review of Particle Properties', \PR {\bf D 50}
(1994), vol. 3, part I and references therein.}. This quantitative
analysis does not confirm the earlier qualitative estimate.}.

Let us first summarize the main features of the formalism \balfor .

\nref\AlPa{G.~Parisi, \PL {\bf B50} (1974) 367;
G.~Parisi, {\it Proc. $11^{th}$ Rencontre de Moriond}, ed. J. Tran Thanh Van,
ed. Fronti\`eres, 1976;
G.~Altarelli and  G.~Parisi, {\it Nucl. Phys.} {\bf B126} (1977) 298 \semi
See also:
Yu.L.~Dokshitser, {\it Sov. Phys. JETP} {\bf 46} (1977) 641 \semi
V.N.~Gribov and  L.N.~Lipatov, {\it Sov. J. Nucl. Phys.}
{\bf 15} (1972) 438; L.N.~Lipatov, {\it Sov. J. Nucl. Phys.} {\bf 20} (1974)
181.}

In the model of \balfor\ is introduced, beyond the usual Altarelli-Parisi
\AlPa\ QCD evolution, the radiation of bound states
(pseudoscalar mesons) by quarks.
This leads to modify  the evolution equations, adding new splitting functions,
functions that describe this phenomenon.

In Ref. \balfor\ the mesons ($\Pi(k)$)  are coupled to quarks by the effective
action
\eqn\seff
{S_{\rm eff}=\int\!d^4p\;\bar\psi (p)i\slash{D}\psi(p)+
\int\!d^4p\int\!d^4k\;\bar\psi(p+\half k)f_{\Pi}
\chi(k,p)U(k)\psi(p-\half k),}
where $U(k)\equiv\exp\big(i\gamma_5\Pi(k)/f_\Pi\big)$,
$f_\Pi$  is the meson decay constant and the coupling is defined by
\eqn\bsdef
{\chi (k,p)\equiv S^{-1}(p+\half k)
\matele{0}{\psi(p-\half k)\bar{\psi}(p+\half k)}{\Pi(k)}S^{-1}
(p-\half k),}
where $S^{-1}(p)=(\pslash
+\Sigma(p^2))^{-1}$ is the quark propagator ($\Sigma (p^2)$
is the self-energy).

The effective coupling $\chi (k,p)$ is then expanded, in the most
general case, in terms of four vertex functions (a pseudoscalar, two
axials and a tensor.)

This permits us to evaluate the cross section
$\sigma^{\gamma^*X}_{q_iq_j}(x;t)$ for the absorption of a virtual photon
$\gamma^*$ and the emission of a pseudoscalar meson $X$
and, then, the splitting function
\eqn\splitX
{\left[{\cal P}_{q_iq_j}(x;t)\right]_X=
\frac{d}{dt}\sigma^{\gamma^*X}_{q_iq_j}(x;t).}

For the sake of simplicity, considering that the mixing angle is small
(see, for example,
\ref\iota{M.~Genovese, D.~B.~Lichtenberg and E.~Predazzi, \ZP {\bf C61}
(1994) 425 and references therein.}), we assume in the following that
$\eta$ and $\eta'$ are the pure octet and singlet respectively.
One obtains for the non-diagonal  (${\cal P}_{qq}^{ND} $, flavour changing) and
for the diagonal (${\cal P}_{qq}^{D} $) splitting functions:
\eqn\lightev
{\eqalign{&{\cal P}_{ud}^{ND} = {\cal P}_{du}^{ND} =
{d\over dt}\sigma^{\gamma^*\pi^+} ={d\over dt}\sigma^{\gamma^*\pi^-} \cr
&{\cal P}_{uu}^D = {\cal P}_{dd}^D={d\over dt}
\left({\half\sigma^{\gamma^*\pi^0}+\smallfrac{1}{6}
\sigma^{\gamma^*\eta}
+\third\sigma^{\gamma^*\eta '}}\right).\cr}
}
The strange mesons ($K$) contributions are kinematically suppressed and
therefore neglected. In the following we will thus neglect anomalous
contributions to the strange sea evolution.

In \BBFG\ the distribution
\eqn\qplus{q^+(x)\equiv u(x)+\bar u(x)- d(x) -\bar d(x)}
was studied.

For this the (mesonic) non-singlet evolution is given by
\eqn\evolqp
{{d\over dt}{(u-d)^+}=
({\cal P}_{qq}^D-{\cal P}_{qq}^{ND})\otimes {(u-d)^+} .}

Because of the strong mass difference between $\eta '$ ($m_{\eta '}
= 958$ MeV/$c^2$) and pions ($m_{\pi^0} = 135$ MeV/$c^2$),
which is caused by $U(N_f)$ breaking due to axial anomaly
\ref\uone{G.~'t~Hooft,
\PRep\vpy{142}{357}{1986} and references therein.}, there is a
relevant difference between ${\cal P}_{qq}^D$ and
${\cal P}_{qq}^{ND} $, which is responsible of quark-antiquark
sea $SU(2)_F$ symmetry breaking.

The fact that the $U(N_f)$ breaking concerns pseudoscalar mesons only
implies that these bound states only contribute to non-singlet evolution:
for other mesons' multiplets,  ${\cal P}_{qq}^D$ and
${\cal P}_{qq}^{ND}$ are almost equal and moreover they are
kinematically suppressed (of about one order of magnitude) due to
the larger masses of these mesons.

The explicit calculation of cross sections
$\sigma^{\gamma^*X}_{q_iq_j}(x;t)$ \balfor\ shows that they depend on
scalar ($\varphi$, which dominates at intermediate $Q^2$) and
axial ($\tilde\varphi$, which determines the $Q^2$ tail)
vertex functions only.

A dipole form is then assumed  for the vertex functions ($p$ is the
quark momentum, ${f_\pi}= 93$ MeV)
\eqn\ffp
{\varphi(p^2)=\frac{m_d}{f_\pi}\frac{\Lambda^2+m_d^2}{\Lambda^2+p^2},}
\eqn\ffa
{\tilde\varphi(p^2)=\frac{g_\pi}{f_\pi}\frac{\tilde\Lambda^2+m_d^2}
{\tilde\Lambda^2+p^2},}
in terms of the three parameters $\Lambda$, ${\tilde \Lambda}$ and
$g_\pi$.
In \balfor\ $\tilde \Lambda = \Lambda$ is hypothesized (the dependence
of the results on the precise value of $\tilde \Lambda$ is small  anyway),
while $g_\pi$ is between 0 and 1.

The $\Lambda$ value may be related, using chiral Ward identities, to
the constituent quark mass. This requires it to vary
in the interval $0.4~$GeV$/c^2~\lsim\Lambda\lsim 0.8~$GeV$/c
^2$.
The $\Lambda$ and $g_\pi$ values have been  fixed in Refs. \refs{\balfor ,
\BBFG } to $\Lambda=0.4 $GeV$/c^2$ and $g_\pi=0.5$
by the request of reproducing the NMC Gottfried sum datum \ref\NMCSG{
P. ~ Amaudruz \etal , \PR {\bf D 50} (1994) R1.}.

In \balfor\ it has also been calculated that the anomalous dimension
due to pion emission is significantly different from zero in the
region 0.05 (GeV/$c)^2 \lsim Q^2 \lsim$ 5--10 (GeV$/c)^2$.
The evolution must therefore be extended up to very low $Q^2$.

The strategy of \BBFG\ was to insert the effect of the meson emission
inside the model of Ref. \ref\glulowq{V.~Barone, M.~Genovese, N.~N.~Nikolaev,
E.~Predazzi and B.~G.~Zakharov, \break
\ZP\vpy{C58}{541}{1993}; \IJMP\vpy{A8}{2779}{1993}.}, which permits an
extension to the $Q^2$ evolution up to very low  $Q^2 $ in order to study the
non-singlet evolution starting from a reasonable ansatz at
$Q^2 \approx 0 ($GeV$/c)^2$.

In the following we will adopt a different strategy in order to study
the anomalous singlet evolution, namely we will begin the evolution at
an input scale $Q^2_0$ sufficiently high to allow other
non-perturbative effects on $Q^2$ evolution to be neglected
a part from the meson emission ones (in the following we will assume as a
reasonable
choice $Q^2 \gsim $1--1.5 (GeV$/c)^2$).
This allows us also to use as input an ordinary parametrization of the
experimental structure functions.

As we are evaluating a first approximation of these effects, we
will neglect (as in \refs{\balfor , \BBFG })
the effects due to mesons other than the pseudoscalar ones,
which are expected to be kinematically suppressed due to larger masses.

For the singlet evolution one has to consider the full set of
coupled equations (omitting gluon contributions for simplicity):
\eqn\appi
{\eqalign{{d\over dt} q_i &=\sum_j  {\cal P}_{q_iq_j}\otimes q_j+
\sum_j  {\cal P}_{q_i\bar q_j}\otimes q_j +
\sum_a {\cal P}_{q_i \Pi^a}\otimes \Pi^a,\cr
{d\over dt} \bar{q}_i &=\sum_j {\cal P}_{\bar{q}_i\bar{q}_j}\otimes \bar{q}_j+
\sum_j {\cal P}_{\bar{q}_i q_j}\otimes q_j+
\sum_a {\cal P}_{\bar{q}_i \Pi^a}\otimes\Pi^a,\cr
{d\over dt} \Pi^a &=\sum_j {\cal P}_{\Pi^a q_j} \otimes q_j +
\sum_j  {\cal P}_{\Pi^a \bar{q}_j} \otimes\bar{q}_j+
\sum_b {\cal P}_{\Pi^a \Pi^b}\otimes\Pi^b,\cr}}
where $\Pi^a(x;t)$ are the distribution functions of
mesons of the pseudoscalar multiplet.

The distributions must fulfill the momentum sum rule
\eqn\msr{\int_0^1 dx \cdot x \cdot \left[ \sum_j \left( q_j + \bar{q}_j \right)
+ \sum_a \Pi_a + g \right]= 1}
where $g$ is the gluon distribution.

The first two equations in
\appi\ express the evolution of the quark and antiquark distribution due to
mesons emission, whereas the last equation gives the evolution of the
pseudoscalar meson distribution.
The former are determined by the splitting
functions ${\cal P}_{q q}$,  which expresses the
probability of  a quark to be emitted by another quark
(with emission of an unobserved  meson), and  ${\cal P}_{q \Pi}$,
which expresses the probability of a pseudoscalar to fragment into a
quark and an antiquark  (one of which is observed); at first order
${\cal P}_{q \bar q}$ is vanishing.
The latter
is determined by the splitting functions ${\cal P}_{\Pi q}$
and ${\cal P}_{\Pi\, \Pi}$, which give the
probability of a pseudoscalar to be emitted by a quark or another pseudoscalar,
respectively.

It must be noticed that not
all these splitting functions are independent, being related by
isospin and charge-conjugation invariance (see \balfor ).

Considering that we will be interested in the medium-to-large $x$ region (far
from the low $x$ region where one expects a rapid growth of the
singlet structure
function due to the radiative generation of the sea from the glue), we will
neglect, as a first approximation,
the effect of the splitting of the radiated mesons in a $q \bar q$ pair,
which is expected to contribute substantially to the low $x$
region only and that
is in any case also suppressed by the fact that the mesons distributions are
not
expected to be very large \BBFG .

We will impose the momentum conservation, adding, as usual \ref\rob{See for
example: R.~G.~Roberts, `The structure of the proton',
Cambridge University Press (1990).},
a delta contribution to the splitting functions:
\eqn\PDeltaD{P^D(z) = {\cal P}^D (z) - {\cal P}_{\Delta} \delta (z-1)}
and
\eqn\PDeltaND{P^{ND} (z) = {\cal P}^{ND}(z) - {\cal P}_{\Delta} \delta (z-1),}
where
\eqn\Pd{{\cal P}_{\Delta}= \int_0^1 dz  \cdot z
\cdot ( {\cal P}^D (z) + {\cal P}^{ND} (z) ).}

The fact that we have neglected the mesonic distributions is thus
phenomenologically compensated,
in order to keep the validity of the momentum sum rule \msr ,
by the explicit choice of the  value of $ {\cal P}_{\Delta}$.

The $F_2$ evolution is then given by:
\eqn\evolv{ \eqalign{{d F_2(x,t) \over dt} =& \int_x^1 {dy \over y} \left[
P_{AP}(x / y) + P^D(x / y) +
P^{ND}(x / y) \right] \times \cr
& \left[ 4/9 \cdot (u(y)+\bar u(y))+  1/9 \cdot (d(y) + \bar d(y))
\right] + \cr &
P_{AP}(x/y) \times 1/9 \cdot \left[s(y) +\bar s(y) \right], \cr }}
where the mesonic effects concern the $u$ and $d$  quarks only (as we
are neglecting the kinematically suppressed contributions of strange mesons)
and $P_{AP}$ are the usual Altarelli-Parisi splitting functions \AlPa .
The heavy quarks, considering the low $Q^2$ under investigation, are completely
negligible.

The numerical calculations are made starting with an input for the
different parton distributions at $Q^2_0=1.6 ($GeV$/c)^2$; in the following
we will show the results obtained using as input the CTQE3M distributions
\ref\CTEQ{H.~L.~Lai \etal, CTEQ-404 (1994).}; however, we have explicitly
checked that our results are largely independent from this choice.

We carry out the evolution \evolv\ through many
small steps in $Q^2$ (we have explicitly verified the stability of our
results on the choice of the steps),
adding at every little step in $Q^2$, the mesonic contribution
to the one coming from the Altarelli-Parisi part (at leading order).
This last part of the evolution is effected in the momentum space (in order to
have a better numerical accuracy); therefore,
at each $Q^2$ step,
the singlet, non-singlets and glue distributions are parametrized in the form
$ p_1 \cdot x^{p_2} \cdot (1-x)^{p_3} \cdot (1+ p_4 x ^{0.5}+ p_5 x +
p_6 x^2 + p_7 x^3)$, which permits
an analytical Mellin transform to the momentum space. The inverse Mellin
transform to $x$ space is made numerically.
The value $\Lambda_{LO}^{nf=3} = 200$ MeV/c is used.

The mesonic effects tend to increase the growth of the parton distributions
at low $x$ (under $x \approx 0.15$), while at larger $x$ the effect of the
delta term dominates producing a more negative ${d \log F_2 \over d t}$.
Altogether this therefore simulates a larger value of $\alpha_s(Q^2)$, except
for a very narrow region around the point where the mesonic contribution
changes
of sign.

In \fig\dffig{The ratio ${d \log F_2 \over d \log Q^2}$
obtained with (solid curve) and
without (dashed curve) mesonic effects at $Q^2=1.6 ($GeV$/c)^2$. The value
$\Lambda_{LO}^{nf=3} = 200$ MeV/c is used.}  ${d \log F_2 \over d t}$
is shown, as obtained with (solid curve) and
without (dashed curve) mesonic effects at $Q^2=1.6 ($GeV$/c)^2$.

\noindent For $x \simeq 0.6$--$0.7$ there is a  difference of
0.03 ($\sim 10 \%$) between the two predictions
(at higher $x$ non-leading twist effects, as target mass corrections,
may be quite important).
At this scale
the mesonic effects simulate an increase of $\Lambda_{LO}^{nf=3}$ from
$200 $ MeV/c to $210$--$220 $ MeV/c. At larger scales the effect reduces
rapidly.
It is already small at $Q^2 \simeq$ 5 (GeV/$c)^2$ (where it is reduced
by about one half of its former value), albeit always
leading to an overestimate of $\Lambda_{QCD}$ of $\approx 5 \%$, and it is
negligible for $Q^2 \simeq$ 10 (GeV/$c)^2$.

It is thus not yet really possible to observe this effect in today's DIS
data; anyway the order of magnitude of the effect at $Q^2 \approx$ 1
(GeV/$c)^2$
is such that it could constitute a
non completely negligible source of error in the determination of $\alpha_s$
by DIS data. A precise determination of this error  obviously depends on the
weights assumed  by different $Q^2$ and $x$ regions in a global fit; however,
one can roughly expect a possible overestimate of $\Lambda_{QCD}$ of a $\approx
5 $--$ 10 \%$ for a global fit in a region such as  $1 ~($GeV$/c)^2~ \lsim Q^2
\lsim 5 ~$(GeV$/c)^2$ and $x \gsim 0.01$.

When higher precision experimental data will become available it would
surely be
quite interesting to search for a confirmation of this effect by comparing
${d \log F_2 \over d t}$ evaluated with $\Lambda_{QCD}$
obtained in other processes
and the one extracted by DIS data in the $Q^2 \lsim 5 ($GeV$/c)^2$ region.
However, a direct observation of this effect could turn out to be
rather difficult due to the fact that scale fixing and other theoretical
uncertainties could lead to an arbitrariness in determining
a precise $\Lambda_{QCD}$ value, which is of the same order as this effect.

The study of the singlet anomalous evolution enables us to evaluate the
anomalous evolution effects on $d (F_2^n / F_2^p) \over d t$ also (where
$F_2^n$ is the neutron structure function and $F_2^p$ the proton one.)

It must be noticed that the numerical estimate $d (F_2^n / F_2^p) \over d t$
is much more sensitive to the choice of the parton distributions than
the evaluation of the derivative of $F_2$. The estimate of the effect of
the mesonic contributions to this quantity is thus, for a precise numerical
calculation, partially depending on this choice as well.
Anyhow for a first evaluation of the effect this dependence is not really
severe, as we have checked with other inputs beyond the
CTEQ3M parton distributions (used as input at $Q^2= 1.6 ($GeV$/c)^2$
as in the $F_2$ derivative case).

 In this case the mesonic effect is a positive contribution to
$d (F_2^n / F_2^p) \over d t$ in the whole range of $x$. It enhances
this quantity with respect to the one evaluated with only  the Altarelli-Parisi
splitting functions  of
$\approx 0.015$ at $Q^2= 1.6 ($GeV$/c)^2$ in the region $0.1 < x < 0.6$.
The effect is still larger at higher $x$, where a careful analysis of
target  mass effects and of other higher twist effects would  anyway
be necessary,
while it disappears at lower $x$ (at $x=0.01$ it is reduced to a difference
of 0.003).

The effect decreases with $Q^2$ and it is reduced by $50\%$ at $Q^2=10
($GeV$/c)^2$.

Nowadays experimental data \ref\NMCr{P.~Amaudruz \etal , \NP {\bf B 371}
(1992) 3.} are not yet able to discern this effect because of large errors.
Anyway they are in qualitative agreement with this prediction for $x \approx
0.4$--$0.6 $, where they show values larger than zero, above QCD predictions,
while they seem to point out more negative values than the QCD prediction
for $ x \approx 0.1$--$0.3$. In \NMCr\ this effect has been
interpreted  in terms of higher twist effects.

More precise experimental data and a more careful treatment of higher twist
effects (for example by fitting them carefully in high precision experimental
data or by evaluating them in some theoretical scheme)
could lead in a next future to an  at least qualitative test of this
prediction.

In summary we have presented an estimate of the effects of the anomalous
evolution due to the mesonic degrees of freedom on singlet structure functions
in the scheme of \balfor .
The effects have been found to be not very large; nevertheless they
are not negligible for
$1 ~($GeV$/c)^2~ \lsim Q^2 \lsim 5 ~($GeV$/c)^2$ and could constitute a
non-negligible source
of error in a high precision determination of $\alpha_s$ from DIS
including data in this region.

\vskip 2cm
{\bf Acknowledgements}
\vskip 1cm

We are grateful to S. Forte and R. Ball for many useful discussions and
comments.
\vfill\eject
\listrefs
\vfill\eject
\listfigs
\bye